\documentclass[aps,prb,twocolumn,superscriptaddress,showpacs]{revtex4-1}
\usepackage{amsmath,amssymb,amstext,graphicx,bm}
\usepackage{graphicx,color,bm}
\usepackage[english]{babel}
\usepackage{dsfont}
\usepackage{epsfig}
\begin{document}

 \title{Killing Auger recombination %processes 
 in nanostructures by carrier spin polarization} 

 \author{Alexander Khaetskii}
\affiliation{Department of Physics, University at Buffalo, SUNY, Buffalo, NY 14260-1500} 

\author{Igor~\v{Z}uti\'{c}}
\affiliation{Department of Physics, University at Buffalo, SUNY, Buffalo, NY 14260-1500}

\date{\today}

%\draft

\begin{abstract}
In semiconductor nanostructures nonradiative Auger recombination is enhanced by the presence of boundaries which relax the momentum conservation and thereby removes the threshold reduction for these processes. We propose a method to strongly reduce the Auger recombination rate by injecting spin-polarized carriers. Our method is illustrated on the example of a quantum well in which the spin-orbit coupling of conduction band is negligible as compared to valence band and thus holes can be considered as spin-unpolarized.  The suppression factor of the Auger recombination is determined by the two-dimensional character of the system, given by the ratio of the Fermi energy of electrons and the separation of the electron levels quantized in the growth direction. Our predictions can be tested experimentally and we discuss their implications for semiconductor lasers relying on injection of spin-polarized electrons.
\end{abstract}

\pacs{     }

\maketitle

%\begin{multicols}{2} 

Auger recombination (AR)  is the dominant nonradiative process in many semiconductors.\cite{Landsberg:1991,Auger}
It is recognized for its limiting role in the performance, not just of lasers, light-emitting diodes,  and solar cells, but also of
transistors and other devices whose performance is governed by lifetimes.\cite{Landsberg:1991}  
In a simple picture, AR 
rate is cubic in a carrier density, unlike the quadratic dependence of the radiative recombination rate.  These trends already suggest an increased role of the AR  with heavy doping and scaling-down of devices. However, even more drastic changes result when bulk semiconductors are replaced by their nanostructures. 

In the AR, the recombination energy of an electron-hole pair is transferred to a third carrier, for example, an electron, as depicted in Fig.~1. In bulk semiconductors the AR  is strongly suppressed at low temperatures.  The simultaneous conservation of the energy and momenta of colliding particles leads to a large energy threshold.\cite{Landsberg:1991}  
 However, in quantum wells, the quantum confinement removes the momentum conservation along the quantization direction.  Colliding particles  can exchange their momenta with the boundaries of the confinement  and the AR becomes thresholdless.\cite{Zegrya,Dyakonov,Polkovnikov}
 
The resulting strong enhancement of AR is not limited to quantum wells (QWs), but 
ubiquitous in many other nanostructures. For example,  in colloidal quantum dots (QDs), AR is attributed to 
intermittent flourescence (blinking), permanent photobleaching, and  multiexciton decay.\cite{Klimov2008:PRB,Hollingsworth2013:CM}
Consequently, there is a great interest to overcome this problem and suppress the AR  rate in
semiconductor nanostructures. For example, using strain was considered for AR suppression in QWs 
(with contradictory results),\cite{Adams1986:EL,Wang1993:APL,Fuchs1993:APL}
while in colloidal  QDs it was suggested that smoothing the confinement potential could
significantly reduce the AR rate, as compared to abruptly terminated boundaries.\cite{Efros}

\begin{figure}
\begin{center}
%\centerline{
\resizebox{4.3in}{!}{\includegraphics{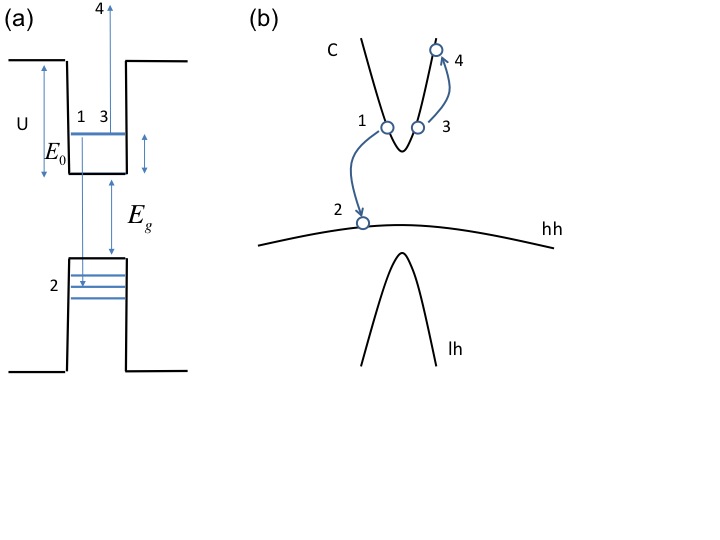}} %}
\vspace{-2.8cm}
\end{center}
\caption{Auger recombination in a quantum well. (a) Energy levels and the quantum confinement. (b) Schematic band structure, 
C: conduction band, hh (lh): heavy (light) hole.
Electron (1) recombines with a heavy hole (2) exciting another electron (3) to the final state (4). } 
\label{fig:1}
\end{figure}

In this work we focus on QWs and  propose a different mechanism for the suppression of the AR rate which relies on 
injecting the spin-polarized carriers. Considering that such a spin injection is one of the key elements for spintronic
devices and many spin-dependent phenomena,\cite{Zutic2004:RMP,Fabian2007:APS,Dyakonov:2008,Maekawa:2006} 
it would be important to understand the role of carrier spin polarization on AR.

With experimental advances in semiconductors spin-laser which rely on injected spin-polarized 
carriers,\cite{Hallstein1997:PRB,Rudolph2003:APL,Holub2007:PRL,Hovel2008:APL,Saha2010:PRB,Gerhardt2011:APL,Iba2011:APL,%
Frougier2015:OE,Cheng2014:NN,Alharthi2015:APL}
the suggested AR suppression may extend the limits of what is feasible in conventional (spin-unpolarized) lasers. For example, in such spin-lasers a steady-state lasing threshold reduction\cite{Rudolph2003:APL,Holub2007:PRL,Gothgen2008:APL,Vurgaftman2008:APL}and spin-filtering,\cite{Iba2011:APL} could also be accompanied with the improved dynamical performance in the modulation bandwidth, switching properties, and eye diagrams,\cite{Wasner2015:APL}
suitable for high-performance optical interconnects.\cite{Faria2015:PRB}

We consider a usually dominant AR in QWs,\cite{Dyakonov} denoted by  CCCH in a conventional notation:\cite{Landsberg:1991}  a recombination of an electron (1) with a heavy hole (2) is accompanied by an  excitation of another electron (3) to the final state (4), 
see Fig. 1. We assume, as it is typically the case,  that  the heavy hole mass is much bigger than the mass of the light hole.  As a result,
the heavy holes density is much bigger than that of the light ones and the electron-heavy hole recombination dominates.
In the initial state the carriers are confined within the QW, their momenta are being defined by the temperature or the Fermi energy,  while the Auger electron (4) is ejected out of the QW with a large momentum directed almost perpendicular to the QW plane.
Due to the presence of the heteroboundaries the momentum is not conserved, the characteristic of the thresholdless processes. 
This non-conservation of the carriers' momenta makes the AR in heterostructures so important and detrimental for lasers and other devices.

In our analysis we focus on the spin-polarized electrons, described by the electron density spin polarization,
\begin{equation}
P_n=\frac{n_{\uparrow}-n_{\downarrow}}{n_{\uparrow}+ n_{\downarrow}}, %\,\,\, n_{\uparrow}+n_{\downarrow}=n.
\label{eta}
\end{equation}
where the total two-dimensional (2D) electron density, $n$,  is the sum of  the spin-up and spin-down 
contributions, $n=n_{\uparrow}+ n_{\downarrow}$, while the holes are considered spin-unpolarized. 
This is readily  realized experimentally by electrical spin injection of electrons, or an optical injection by circularly polarized 
light in which the holes loose their spin polarization nearly instantaneously 
(for example, 3-4 orders of magnitude faster than electrons in GaAs at room temperature), as a consequence of much stronger
spin-orbit coupling in the valence band.\cite{Zutic2004:RMP,Dyakonov:2008} Equivalently, we expect that the hole spin-relaxation time
is shorter than the AR time. Assuming spin-unpolarized holes is also widely used in the studies of 
spin-lasers.\cite{Rudolph2003:APL,Holub2007:PRL,Hovel2008:APL,Gothgen2008:APL,Vurgaftman2008:APL}

Formally, we consider AR in materials with  spin-orbit splitting of the valence band much smaller than the band gap,
$\Delta \ll E_g$, and neglect the spin-orbit coupling in the conduction band. The electrons occupy only the first quantum level in the well, i.e.  $k_B T < \epsilon_F\ll \delta E_z$, where $\delta E_z$  is the energy separation between the quantized levels in growth direction, $T$ is the temperature, and $\epsilon_F$ is the Fermi energy of electrons.
\par
In order to understand the physical picture, let us consider the limiting case when the electrons are fully spin-polarized. Since two electrons (1) and (3) in the initial state belong to the same quantum level in z-direction and differ only by the values of the momenta within the 
plane, it is clear that 
in the lowest approximation with respect to the parameter  $\epsilon_F/\delta E_z \ll 1$ their quantum states are the same and the symmetrized wave function of two-particle state is zero (suppression is due to the Pauli principle).   It means that non-zero contribution to the AR rate appears only as a square of the  difference between the longitudinal momenta of the initial electrons. Thus, the rate  will be proportional to the above mentioned parameter. This small parameter describes the AR rate suppression 
for the spin-polarized initial state, as compared to the rate for the unpolarized case. 
The latter was calculated in Refs.~\onlinecite{Dyakonov,Polkovnikov}, but  answers obtained in these 
papers are mutually different in some limiting cases.  Note that  a more pronounced suppression of the Auger rate occurs for a smaller QW width. 

\par
It is straightforward to obtain the dependence of the AR rate on $P_n$.
The Auger rate is given by the expression  
\begin{equation}
R =\frac{2\pi}{\hbar}\sum_{1,2,3,4}|M|^2\delta(E_1+E_3-E_2-E_4)f_1^cf_3^cf_2^v(1-f_4^c).
\label{rate}
\end{equation}
Here $M$ is the matrix element of the Coulomb interaction, and the sum is taken over all the quantum numbers of the particles, spin included, while
%Functions 
$f^c, f^v$ are the Fermi functions for electrons and holes.  
\par
In the absence of the spin-orbit coupling, the sum over the spins in Eq. (\ref{rate}) may be written for the unpolarized electrons 
as,\cite{Dyakonov} 
%(see Ref. \onlinecite{Dyakonov})
\begin{equation}
\sum_{(\sigma)}|M|^2=\frac{e^4}{\kappa^2}(|M_I+M_{II}|^2 +3|M_I-M_{II}|^2), 
\label{matrix}
\end{equation}
where $e$ is the electron charge, $\kappa$ is a dielectric constant, and 
\begin{eqnarray}
M_I=\int d^3rd^3r' \Psi^{\star}_1(\mathbf{r})\Psi^{\star}_3(\mathbf{r'})\frac{1}{|\mathbf{r}-\mathbf{r}'|}\Psi_2(\mathbf{r})\Psi_4(\mathbf{r'}), 
\label{dir}\\
M_{II}=\int d^3rd^3r' \Psi^{\star}_1(\mathbf{r'})\Psi^{\star}_3(\mathbf{r})\frac{1}{|\mathbf{r}-\mathbf{r}'|}\Psi_2(\mathbf{r})\Psi_4(\mathbf{r'}).
\label{exch}
\end{eqnarray}
Here $M_I$ and  $M_{II}$ are the direct and exchange Coulomb matrix elements, correspondingly. The two terms in Eq. (\ref{matrix}) are due to contributions from the singlet and triplet states of the initial electrons, correspondingly. 
\par
Under the conditions considered in Ref. \onlinecite{Dyakonov} [when the initial momenta of electrons within the 2D plane were small and neglected in Eqs. (\ref{dir}) and (\ref{exch})], 
the matrix elements   $M_I$ and  $M_{II}$  were equal,  and only the contribution which was taken into account was that from the singlet state, i.e. the first term in Eq.~(\ref{matrix}). It reads 
\begin{equation}
\sum_{(\sigma)}|M|^2=\frac{4e^4}{\kappa^2}|M_I|^2 . 
\label{unpolarized}
\end{equation}
When  electrons are spin-unpolarized, this contribution is much bigger than that due to the second term in Eq.~(\ref{matrix}). 
Within the approximation outlined above the initial carrier energies in the argument of the $\delta$-function in Eq.~(\ref{rate}) 
can be neglected (these energies are small compared to $E_g$). Moreover,   
since $|M_I|^2$ does not depend anymore on  the in-plane momenta of the initial electrons $\bf{p}_1, {\bf p}_3$, we can easily perform the summations \cite{Dyakonov} over these momenta in Eq.~(\ref{rate})
\begin{equation}
\sum_{{\bf p}_1,{\bf p}_3,(\sigma)}|M|^2 f_1^c f_3^c=\frac{e^4}{\kappa^2}|M_I|^2 n^,
\label{unpolar}
\end{equation}
here we have used that  $\sum_{{\bf p}_1} f_1^c =\sum_{{\bf p}_3} f_3^c=n/2$.

\par
In the case of an arbitrary spin polarization of the electron system the summation over spins  leads to the expression 
\begin{eqnarray}
\sum_{(\sigma)}|M|^2 f_1^c f_3^c =\frac{e^4}{\kappa^2}[|M_I-M_{II}|^2 (f_{1,\uparrow}^c f_{3,\uparrow}^c+ f_{1,\downarrow}^c f_{3,\downarrow}^c) &+& \nonumber \\
+(|M_I|^2+|M_{II}|^2)(f_{1,\uparrow}^c f_{3,\downarrow}^c +f_{1,\downarrow}^c f_{3,\uparrow}^c)].  
\label{matrix1}
\end{eqnarray}
If we neglect in the expressions for $M_I, \, M_{II}$ the initial in-plane electron momenta,  making $M_I = M_{II}$, 
we obtain 
\begin{equation}
\sum_{{\bf p}_1,{\bf p}_3,(\sigma)}|M|^2 f_1^c f_3^c=\frac{e^4}{\kappa^2}4|M_I|^2 n_{\uparrow}n_{\downarrow}=
\frac{e^4}{\kappa^2}|M_I|^2 n^2(1-P_n^2).  
\label{arbitrary}
\end{equation}
From the comparison of Eqs. (\ref{unpolar}) and (\ref{arbitrary}) we conclude that the AR rate for an arbitrary $P_n$ %spin polarization
is related to the corresponding rate for an unpolarized system through the %following 
simple formula,
\begin{equation}
R(P_n)=R(0)(1-P_n^2).
\label{answer}
\end{equation}
We note that the rate $R(0)$ can include all the processes considered in Ref. \onlinecite{Polkovnikov}, i.e. thresholdless, quasithreshold and threshold ones. 
As it follows from Eq.(\ref{answer}),  for the system of initially fully spin-polarized electrons  the AR rate is totally suppressed. 
In fact, this conclusion is only correct in the lowest approximation used above, when in calculations of $M_I$ and $M_{II}$ the initial in-plane electron momenta ${\bf p}_1,{\bf p}_3$ were entirely neglected. 
Let us estimate the residual value of the AR rate for a fully spin-polarized system, that is the value of $|M_I-M_{II}|^2$, see 
Eq.~(\ref{matrix1}).
From the expressions for the wave functions of electrons and heavy holes presented in Ref.~\onlinecite{Polkovnikov} (we use them at zero spin-orbit coupling) we obtain $|M_I-M_{II}|^2 \simeq  |M_I|^2  ({\bf p}_1-{\bf p}_3)^2/p_0^2$, where $p_0$ is the characteristic electron momentum in the growth direction ($p_0=\hbar\pi/a$ for the square well of the width $a$).  It means that the AR rate for the fully polarized electrons will be suppressed compared to its value  at  zero polarization in accordance with the small parameter $\epsilon_F/E_0 \ll 1$, where $\epsilon_F=\pi \hbar^2 n/m_e$ is the electron Fermi energy  at zero polarization and  $E_0=p_0^2/2m_e$ is the energy of the size quantization in the z-direction. 
We assume that $E_0 \ll U$, where $U$ is the height  of the barrier for the electrons. 
Thus, the estimate for the residual value of the AR rate for a fully spin-polarized system is 
\begin{equation}
R(P_n =1) \simeq R(0) \frac{\epsilon_F}{E_0}. 
\label{answer1}
\end{equation}
If we also take into account a small spin-orbit coupling for the electron states, we obtain that the residual value of the Auger rate for a strongly polarized electron system  will be determined by a small factor $(\Delta/E_g)^2 \ll 1$.
This factor replaces $\epsilon_F/E_0$ in Eq.~(\ref{answer1}).

\par

In conclusion, we have derived a simple formula which describes the supression of CCCH Auger recombination rate in quantum wells with the spin polarization of electron density.  The parameter which describes this suppression is the ratio of the electron Fermi energy and the electron quantization energy in the growth direction. Our prediction can be tested experimentally and further motivates efforts to realize effective spin injection in lasers. In particular, the suppression of Auger recombination with carrier spin polarization provides opportunities for additional advantages of spin-lasers over their conventional counterparts.

This work was supported by NSF ECCS-1508873 and U.S. ONR N000141310754.

\end{document}